# Energy Management Strategy for Unmanned Tracked Vehicles Based on Local Speed Planning

Tianxing Sun, Shaohang Xu, Zirui Li, Yingqi Tan, Huiyan Chen

*Abstract*—The hybrid electric system has good potential for unmanned tracked vehicles due to its excellent power and economy. Due to unmanned tracked vehicles have no traditional driving devices, and the driving cycle is uncertain, it brings new challenges to conventional energy management strategies. This paper proposes a novel energy management strategy for unmanned tracked vehicles based on local speed planning. The contributions are threefold. Firstly, a local speed planning algorithm is adopted for the input of driving cycle prediction to avoid the dependence of traditional vehicles on driver's operation. Secondly, a prediction model based on Convolutional Neural Networks and Long Short-Term Memory (CNN-LSTM) is proposed, which is used to process both the planned and the historical velocity series to improve the prediction accuracy. Finally, based on the prediction results, the model predictive control algorithm is used to realize the real-time optimization of energy management. The validity of the method is verified by simulation using collected data from actual field experiments of our unmanned tracked vehicle. Compared with multi-step neural networks, the prediction model based on CNN-LSTM improves the prediction accuracy by 20%. Compared with the traditional regular energy management strategy, the energy management strategy based on model predictive control reduces fuel consumption by 7%.

## I. INTRODUCTION

As one of the core technologies of hybrid vehicles, energy management links the fuel and electric drive systems. Driving cycle prediction is the key to energy management strategy, and the accuracy of driving cycle prediction directly affects the optimization effect of energy management.

At present, the research on energy management strategies of hybrid electric vehicles is mainly divided into two categories: rule-based energy management strategy and optimization-based energy management strategy [1]. The rule-based energy management strategy has the advantages of a simple algorithm, easy implementation, and high real-time performance, maintaining power battery SOC stability and improving fuel economy [2], [3]. It is a widely used energy management strategy at present. However, the control rules depend on engineering experience, and the control parameters cannot be adjusted in real time [4]. For different driving systems and complex working conditions, rules need to be re-established, and adaptability is lacking [5]. To further improve the fuel economy performance of the hybrid power system, most researches mainly focus on optimization-based energy management strategies [6]. The representative algorithms include dynamic programming, equivalent consumption minimization strategy, model predictive control and reinforcement learning [7]. [8] adopted a dynamic programming algorithm to solve the optimal control problem for plug-in hybrid vehicles, which avoided frequent shifting and reduced fuel consumption. Dynamic programming algorithm can obtain the global optimal control results, but it needs to know the driving conditions in advance. In response to the shortcomings of global optimization strategies, researchers have carried out real-time optimization energy management strategies. [9] adopted an energy management strategy based on model predictive control, compared with the regular energy management strategy, this strategy has better fuel economy performance and a more balanced state of charge. The model predictive control strategy considers the state information of the vehicle at the present moment and the information of the driving cycle prediction at the future moment. Compared with the static energy management strategy, it has better applicability and energy-saving effect. With the development of modern technology, the energy management strategy of the hybrid electric vehicle is becoming more and more intelligent. [10] used a reinforcement learning algorithm to solve the optimal control problem for series hybrid tracked vehicles, reducing fuel consumption and improving calculation efficiency.

The above research on energy management can provide a reference for unmanned vehicles. In the traditional energy management strategy, it is usually necessary to identify the demand power according to the operating instructions, such as the accelerator pedal opening. For unmanned vehicles, it is necessary to consider how to formulate an energy management strategy without driver operation instructions. The result of the speed planning of unmanned vehicles represents the expected driving velocity of the vehicle in the future. Intuitively, the result of speed planning is the driving intention of unmanned vehicles. However, the control system will produce control errors when carrying out the planned trajectory, directly affecting the accuracy of the driving cycle prediction. [11] - [16] used historical data to model driver behavior to predict and identify driving behavior. Using these methods to design a prediction model can reduce the error of directly using the speed planning results.

We previously proposed an energy management strategy for unmanned tracked vehicles based on the global path [17]. This strategy can maintain the stability of the battery SOC and reduce fuel consumption, but it cannot guarantee the real-time optimization of energy management. This paper

Tianxing Sun, Shaohang Xu, Yingqi Tan and Huiyan Chen are with the School of Mechanical Engineering, Beijing Institute of Technology, Beijing 100081, China (e-mail: 3120190386@bit.edu.cn; shaohang_xu@163.com; tyq@bgy.edu.cn; chen_h_y@263.net).
Zirui Li is with the School of Mechanical Engineering, Beijing Institute of Technology, Beijing, 100081 China and Department of Transport and Planning, Faculty of Civil Engineering and Geosciences, Delft University of Technology, Stevinweg 1, 2628 CN Delft, The Netherlands (e-mail: 3120195255@bit.edu.cn).
(Corresponding authors is H. Chen).

proposes an energy management strategy based on local speed planning. Compared with traditional algorithms, this algorithm has three different characteristics:

- The planned velocity sequences and historical velocity sequences are used for the input of the driving cycle prediction to adapt to the unmanned driving mode.
- A prediction model based on CNN-LSTM is proposed, which can significantly improve the prediction accuracy.
- Using the model predictive control algorithm for energy management helps maintain the stability of battery SOC and control the engine operating point in the high-efficiency area, which has a good engineering application value.

The structure of this paper is shown in Figure 1. In Section II, the prediction model based on CNN-LSTM is proposed to improve the prediction accuracy. In Section III, model predictive control is used to optimize energy allocation and reduce fuel consumption. In Section IV, we collected various driving data of the unmanned vehicle and verified the effectiveness of the algorithm through experiments. Finally, in Section V, we draw relevant conclusions and prospect some extensions of work.

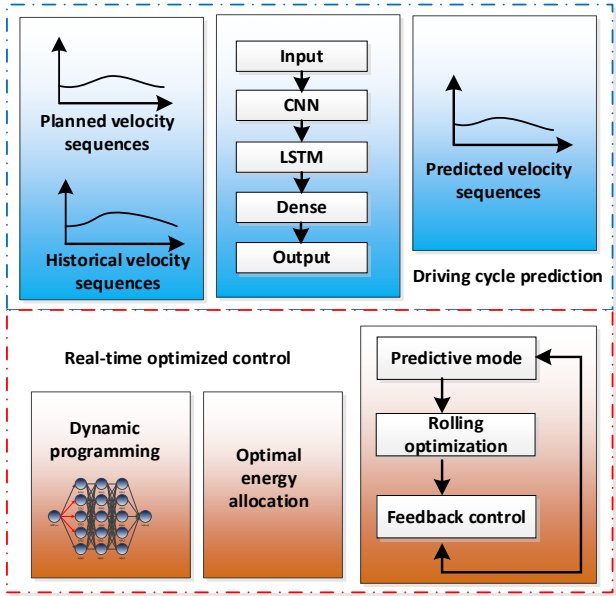

Figure 1.  Energy management strategy based on local speed planning

## II. THE TECHNOLOGY OF DRIVING CYCLE PREDICTION

### A. The Method of Local Speed Planning

The algorithm of local speed planning is detailed in Fig.2 [18]. Firstly, the maximum velocity limit is assigned to the velocity sequences to be calculated. The maximum velocity limit is set in advance according to the driving environment and vehicle dynamic characteristics. Then, the lateral acceleration constraint is applied to the velocity of each path point.

$$v_i \leq \min\left\{\sqrt{a_{max}^{lat}/\kappa}, v_{max}\right\} \quad (1)$$

where $v_i$ is the velocity value of the $i^{th}$ path point, $a_{max}^{lat}$ is the maximum lateral acceleration, $\kappa$ is the curvature of the current path point, and $v_{max}$ is the maximum velocity.

Next, longitudinal acceleration constraints are applied to the velocity of each path point:

$$-d_{max}^{lon} \leq \frac{v_i^2 - v_{i-1}^2}{2|s_i - s_{i-1}|} \leq a_{max}^{lon} \quad (2)$$

where $d_{max}^{lat}$ is the maximum longitudinal deceleration, $a_{max}^{lat}$ is the maximum longitudinal acceleration, and $s_i$ is the arc length from the $i^{th}$ path point to the initial point.

Finally, the jerk of constraint is applied to the velocity of each path point:

$$|j_i| \leq j_{max}^{lon} \quad (3)$$

where $j_i$ is the jerk and $j_{max}^{lon}$ is the maximum jerk.

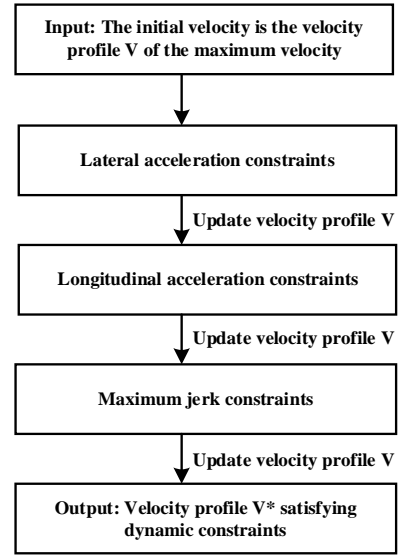

Figure 2.  Velocity profile satisfying system constraints

The velocity of the current path point needs to be adjusted to meet each constraint. The algorithm iteratively modifies the velocity profile until the velocity profile no longer changes, and the final velocity profile is the planned velocity sequence.

### B. Prediction Model Based on CNN-LSTM

In this paper, a model architecture of CNN-LSTM is used to process both the planned velocity sequence and the historical velocity sequence. CNN is used to learn local features trend of time series data；LSTM to obtain long-term relevant features from corresponding time-series data. The algorithm framework is shown in Fig. 3. Firstly, the historical velocity series and the planned velocity series are combined as the input of the CNN. The predicted velocity series is output after the processing of the CNN and LSTM.

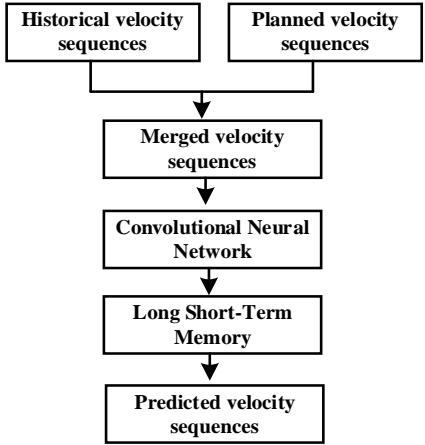

Figure 3. CNN-LSTM model based on planned velocity sequences and historical velocity sequences

## III. REAL-TIME OPTIMIZATION OF ENERGY MANAGEMENT

The purpose of this section is to use a model predictive control algorithm for energy management based on the driving cycle prediction of unmanned vehicles proposed in section II to realize real-time optimization of energy management and improve the fuel economy of vehicles. Firstly, the optimal control problem of energy management is constructed with fuel economy and battery SOC stability as performance indexes, initial conditions and constraints are given. Secondly, the specific process of solving the optimal control problem of energy management using a dynamic programming algorithm is given. Finally, a model predictive control framework based on a dynamic programming algorithm is constructed to realize real-time energy management optimization.

### A. Optimal Control Problem for Energy Management

The optimal control problem of energy management is to solve the optimal control quantity of the performance index under satisfying various constraints of the hybrid power system. The performance function considers the fuel economy of the hybrid power system and the stability of the battery SOC. The performance function in the optimal control problem of energy management is:

$$J = \int_{t_0}^{t_f} \omega_1 \dot{m}(t) + \omega_2 (SOC(t) - SOC_{target})^2 dt \quad (4)$$

where $\dot{m}(t)$ is the fuel consumption rate, $SOC(t)$ is the current battery SOC, $SOC_{target}$ is the target battery SOC, $\omega_1$ and $\omega_2$ are weight factors, respectively.

The SOC of the power battery pack is taken as the state variable of the system, and the state equation is:

$$\dot{SOC}(t) = -\frac{V_{oc}(t) - \sqrt{V_{oc}(t)^2 - 4R_b(t)P_b(t)}}{2R_b(t)C_b} \quad (5)$$

where $V_{oc}$ is the open-circuit voltage of the power battery pack, $R_b$ is the internal resistance of the battery, $P_b$ is the battery power, and $C_b$ is the rated capacity of the battery. Equation (5) is written in the discrete-time form:

$$SOC(k+1) = SOC(k) - \frac{V_{oc}(k) - \sqrt{V_{oc}(k)^2 - 4R_b(k)P_b(k)}}{2R_b(k)C_b} \quad (6)$$

The unmanned tracked vehicle in this paper used a series hybrid power system. The structure of the power system is shown in Fig. 4. The engine output shaft and generator input shaft are directly connected, the engine adopts speed control mode, and the generator adopts torque control mode. The engine speed and generator torque should conform to the external characteristics of the engine and generator, and the change of engine speed and generator torque should not be too fast in the adjacent two control range. Otherwise, the working point transfer cannot be completed, and the engine generator set cannot reach the target output power.

The state variables and control variables need to be limited due to the capability of the components. To prevent overcharge and over-discharge of power battery, the SOC of the power battery pack should fluctuate between the minimum and maximum allowable values, and the battery power should be between the maximum charging power and the maximum generating power. To maintain the stability of the power battery SOC, the initial SOC value of the system remains the same as the end SOC value.

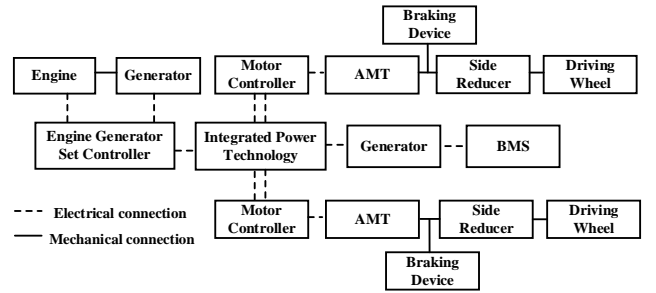

Figure 4. The power system of the unmanned tracked vehicle

### B. The Solution of the Optimal Control Problem

The battery SOC is regarded as the state variable of the system, and the SOC state is discretized. The SOC range is divided into $m$ parts, and its upper and lower limits should meet the requirements of energy management control objectives. The SOC of the battery is $0.6 \sim 0.8$, $n$ is the total number of discrete stages, $k$ is the current stage, $k = 1, 2, 3 \cdots n-1$.

In this paper, a forward dynamic programming algorithm is used to solve the optimal control problem of energy management. For the sampling time $k = 1$ and the $i^{th}$ SOC state $x_i$, there are:

$$J_1(x_i) = L_1(x_i) \quad (7)$$

For the sampling time $k = 2$ and the $i^{th}$ SOC state $x_i$, there are:

$$J_2(x_i) = \min_j \left[ J_1(x_j) + L_2(x_i, x_j) \right] \tag{8}$$

where $J_1(x_j)$ is the minimum cost function of the state $x_j$ at the first sampling moment, and $L_2(x_i, x_j)$ is the instantaneous cost function of the state transition from $x_j$ to $x_i$ under the system constraint in the second stage. Similarly, the minimum cost function of all subsequent stages can be obtained:

$$J_{k+1}(x_i) = \min_j \left[ J_k(x_j) + L_{k+1}(x_i, x_j) \right] \tag{9}$$

where $J_k(x_j)$ is the minimum cost function of $x_j$ state in the $k$ stage, $J_{k+1}(x_i)$ is the minimum cost function in the $k+1$ stage, and $L_{k+1}(x_i, x_j)$ is the instantaneous cost function of state transition from $x_j$ to $x_i$ under the system constraint of $k+1$ stage.

In this way, the minimum cost function is calculated forward until the end time, and the final minimum cost of the optimal control problem can be obtained, that is, the minimum fuel consumption value. According to the minimum cost corresponding to the transfer state of storage, the reverse search of the optimal state can obtain the optimal state trajectory, the SOC value at each moment when the fuel consumption is the minimum and the corresponding control sequence.

At the sampling time $k+1$, the instantaneous cost function of the system is:

$$L_{k+1}(SOC(j), u(j)) = \min \dot{m}_f(u(j)) \tag{10}$$

The calculation of the instantaneous cost function depends on the system state variable $SOC(j)$ and control variable $u(j)$. The system control variables are engine speed and generator torque. The battery SOC and engine speed are selected to determine the state of the system to calculate the instantaneous cost function.

The battery SOC range and engine speed range are discretized. For each SOC discrete value, the SOC change value $\Delta SOC$ at time $k$ to $k+1$ can be obtained:

$$\Delta SOC = SOC_{k+1} - SOC_k \tag{11}$$

From the battery model, the charge and discharge power of the battery in this process can be calculated:

$$P_b = -V_{oc} C_b \Delta SOC - (C_b \Delta SOC)^2 R_b \tag{12}$$

The vehicle demand power can be calculated from known operating conditions:

$$P_{req} = \frac{\left( fmg + \dfrac{CAv^2}{21.15} + ma + mg\sin\theta \right)\dfrac{v}{3.6} + \dfrac{1}{4}\mu mg\omega L}{\eta_m^s \eta_t} \tag{13}$$

where $P_{req}$ is the vehicle power demand, $W$; $f$ is the ground resistance coefficient; $g$ is the gravity acceleration, $m/s^2$; $\theta$ is the ground slope angle; $C$ is the air resistance coefficient; $A$ is the windward area, $m^2$; $v$ is the vehicle speed, $m/s$; $\mu$ is the steering resistance coefficient; $L$ is the track grounding length, $m$; $\omega$ is the rotational angular speed of the tracked vehicle, $rad/s$; $\eta_m^s$ is the efficiency of the motor, $s$ is the charge-discharge coefficient of the motor, $\eta_t$ is the efficiency of the mechanical transmission system.

According to the power balance, the target output power $P_g$ of the engine-generator set can be calculated:

$$P_g = P_{req} - P_b \tag{14}$$

Because of the rigid connection between the engine output shaft and the generator input shaft, the speed, torque, and output power of the engine and generator are equal as follows:

$$\begin{cases} n_e = n_g \\ T_e - T_g = \dfrac{\pi}{30}(J_e + J_g) \\ P_g = \dfrac{T_g n_g}{9.55}\eta_g \end{cases} \tag{15}$$

where $n_g$ and $T_g$ are the speed and torque of the generator input shaft respectively; $J_e$ and $J_g$ are the moments of inertia of the engine and generator respectively; $P_g$ is the electric power output from the generator to the DC bus; $\eta_g$ is the generator efficiency.

The generator torque and engine torque can be obtained by substituting the engine speed into equation (15). Fuel consumption is then obtained from the engine fuel consumption model (equation 16).

$$\dot{m}(t) = \varphi_{fuel}(T_e - n_e) \tag{16}$$

where $\varphi_{fuel}$ is a table lookup function of engine fuel consumption characteristic graph determined according to bench test data; $T_e$ is engine torque; $n_e$ is engine speed.

According to the dynamic programming algorithm, the current total cost is the minimum value of the sum of the total cost corresponding to the state at the last moment and the minimum cost of the transition state, namely:

$$J_{k+1}(SOC(j)) = \min[J_k(SOC(i)) + L_{k+1}(SOC(j), u(j))] \tag{17}$$

The total cost function of the whole process, the optimal control quantity and the state quantity at each moment are obtained until the final termination moment. The energy management strategy process based on model predictive control is shown in Fig. 5.

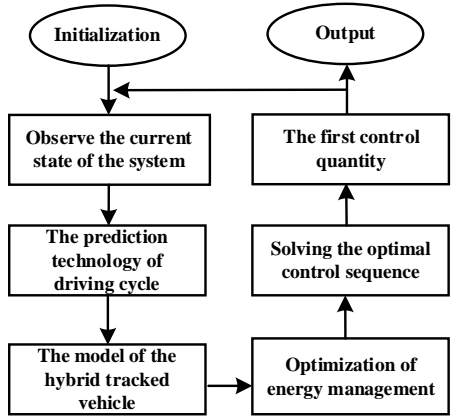

Figure 5. Flow chart of an energy management algorithm

## IV. SIMULATION AND RESULT ANALYSIS

### A. Simulation Experiment and Comparative Analysis Based on CNN-LSTM Prediction Model

The unmanned tracked vehicle used in this paper is shown in Figure 6. The main parameters of the vehicle are shown in Table I. The vehicle has good dynamic performance, complete perception, planning and motion control system, and can autonomously complete environmental detection, obstacle avoidance, overtaking and other unmanned tasks. The test scenario of a tracked vehicle is shown in Fig. 7. In the early stage, many driving data of unmanned tracked vehicles were collected for the training model.

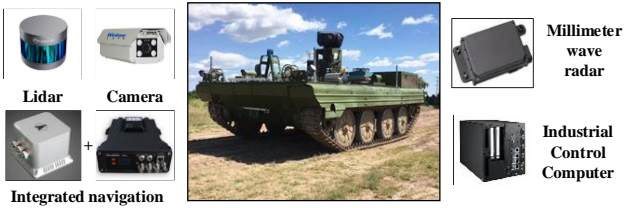

Figure 6. The unmanned tracked vehicle

TABLE I. MAIN PARAMETERS OF THE UNMANNED TRACKED VEHICLE

| Parameter | Values |
|---|---|
| Total quality | 9359kg |
| Transmission ratio | 14.89 |
| Drive wheel radius | 0.2654m |
| Maximum engine power | 96kW |
| Maximum engine speed | 3200rpm |
| Maximum torque of engine | 310Nm |
| Generator rated power | 60kW |
| Generator rated Speed | 2000rpm |
| Generator rated torque | 290Nm |
| Battery capacity | 96Ah |

Fig. 8 shows the prediction results of the neural network model, in which the blue curve is the actual velocity curve, and the red curve is the predicted velocity in each prediction time domain. A three-layer neural network structure is adopted, including a hidden layer with ten nodes. L-BFGS algorithm is used to complete the training and testing of the neural network. To improve the prediction accuracy of the neural network model, upper and lower bound constraints are set for the predicted vehicle velocity in each time domain.

$$\begin{cases} v_p = v_{\max}, v_p > v_{\max} \\ v_p = v_{\min}, v_p < v_{\min} \end{cases} \quad (18)$$

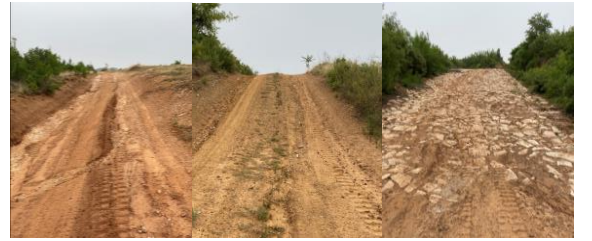

Figure 7. Test scenarios

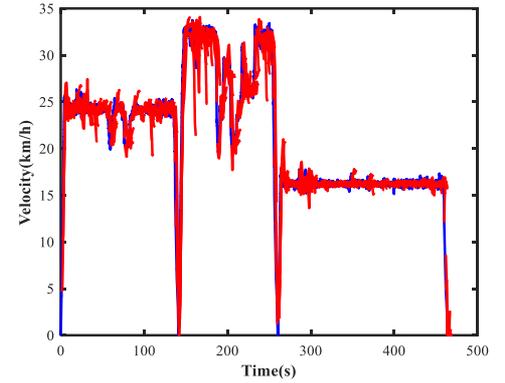

Figure 8. Prediction results of the multistep neural network model

Fig. 9 shows the predicted results of the CNN-LSTM model. The prediction accuracy exceeds the multi-step neural network.

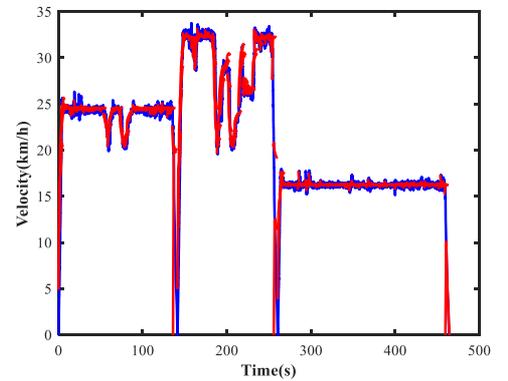

Figure 9. Prediction results of CNN-LSTM model based on planned velocity and historical velocity

Table II shows the comparison of prediction accuracy of different models, in which the mean square error is used as the evaluation index for prediction accuracy, and the calculation formula of root mean square error is shown in Equation (19):

$$\begin{cases} e = \dfrac{\sum_{k=1}^{n} e(k)}{n} \\ e(k) = \sqrt{\dfrac{\sum_{i=1}^{p}(v_p(k+i) - v_r(k+i))}{p}} \end{cases} \quad (19)$$

where $e$ is the root mean square error; $v_p$ and $v_r$ are the predicted velocity and the actual velocity respectively.

TABLE II. COMPARISON OF PREDICTION ACCURACY

| Prediction method | RMSE (km/h) |
|---|---|
| Index prediction model ($\theta=0.05$) | 0.1774 |
| Markov model | 0.1934 |
| Multi-step neural network prediction model with upper and lower bounds corrected | 0.1288 |
| predictive model based on planned velocity | 0.1123 |
| CNN-LSTM model based on planning velocities and historical velocities | 0.1021 |

The result of index prediction is static, unable to reflect the dynamic characteristics of vehicle velocity, and there will be a relatively large prediction error in acceleration and deceleration. The state quantity of the Markov model is relatively simple, and the velocity prediction cannot satisfy the driving characteristics of vehicles well. The multi-step neural network considers the current state and considers the prediction state of the previous step in the multi-step prediction, so it achieves a good prediction accuracy. The model based on CNN-LSTM inherits the characteristics of the historical velocity and adds the unique velocity planning information of the unmanned driving system. The prediction accuracy is more than 20.7% compared with the neural network. The experiment shows that the prediction effect can be improved by using the planned velocity and the historical velocity simultaneously, and the validity of the CNN-LSTM model is verified.

### B. Comparative Verification of Energy Management

To further verify the effectiveness of the energy management strategy proposed in this paper, the energy management strategy based on power following, the prediction strategy based on neural network, and the energy management strategy based on local speed planning proposed in this paper are compared and analyzed in the same simulation environment. The simulation working condition is shown in Figure 10. The simulation sampling time is 1 second, and the prediction time domain and control time domain is 5 seconds.

Fig. 11 shows the change of battery SOC under different strategies. During the period from the 50s to 150s, the driving condition of the tracked vehicle is relatively stable, the power demand is relatively stable, and the battery SOC is stable at the target value. During the 150s to 180s, the tracked vehicle demands a large amount of power, and the battery and engine jointly provide electric energy. During the 180s to 210s, the power demand of the tracked vehicle is small. The output power of the engine-generator set is not only used to drive the vehicle but also to charge the battery, making the battery SOC rise to the initial value. It can be seen from the figure that all three energy management strategies can maintain the battery SOC around the target value. However, the battery SOC of the energy management strategy based on power following has relatively tiny fluctuations, and the engine operating points are not effectively adjusted through the battery. The other two energy management strategies fluctuate significantly, which effectively make the battery play the role of peak clipping and valley filling to adjust the working point of the engine to work more in the high-efficiency area.

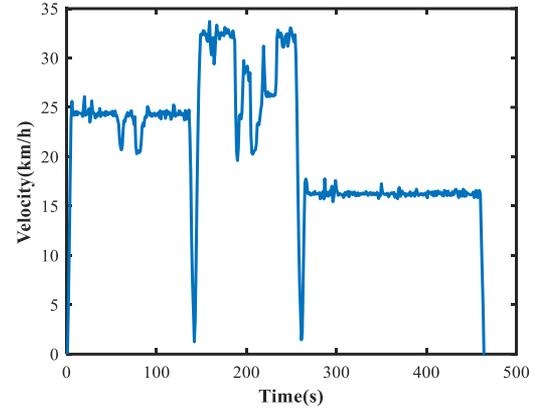

Figure 10. The simulation conditions

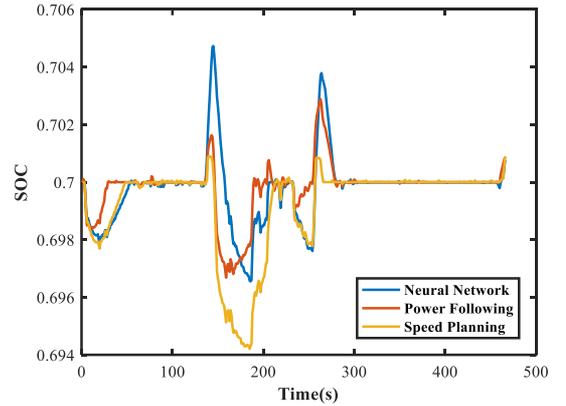

Figure 11. Simulation results of battery SOC for three energy management strategies.

Fig. 12 shows the comparison of simulation results of engine operating points under three energy management strategies. The strategy based on neural network prediction also uses the model predictive control framework proposed in this paper, and the prediction method uses the neural network prediction model. It can be seen that the power following strategy only determines the control amount according to the current error, so the fluctuation of the engine operating point is relatively large. More working points are located in the high-efficiency area with the local speed planning strategy, and the operating range is improved. After the SOC correction, the comparison of equivalent fuel consumption of the three energy management strategies is shown in TableIII. The fuel economy of the two energy management strategies based on MPC is higher than that of the power following strategy, increasing by 4.1% and 7.3%, respectively, which indicates that the energy management based on MPC can improve the fuel economy. Moreover, compared with the neural network prediction method, the driving cycle prediction method proposed in this paper has a higher fuel economy.

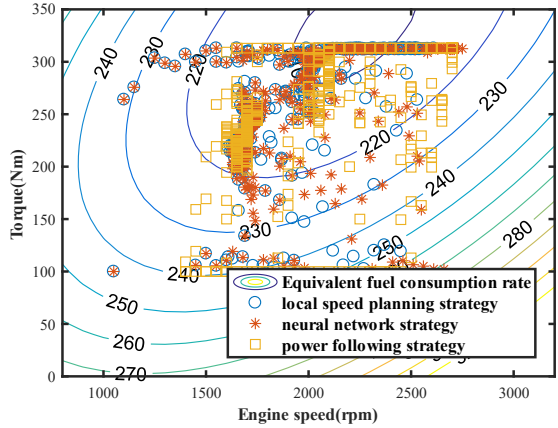

Figure 12. Comparison of engine operating point simulation results of three energy management strategies

TABLE III. COMPARISON OF EQUIVALENT FUEL CONSUMPTION OF THREE ENERGY MANAGEMENT STRATEGIES

| Energy management strategy | Equivalent fuel consumption (g) | Improvement |
| --- | --- | --- |
| Rule strategy based on power following | 1470.1 | - |
| Prediction strategy based on neural network | 1409.7 | 4.1% |
| Prediction strategy based on local speed planning | 1362.8 | 7.3% |

## V. CONCLUSIONS

Due to the complexity of the driving environment and the uncertainty of the driving cycle of unmanned tracked vehicles, this paper proposes a method of driving cycle prediction based on local speed planning. Aiming at predicting error between planned speed and actual speed, a prediction model based on CNN-LSTM is proposed to improve the prediction effect. Based on driving cycle prediction, this paper proposes an energy management strategy based on model predictive control. The simulation results show that the strategy can reduce fuel consumption and improve fuel economy.

We plan to expand on this article in some areas. Firstly, the driving cycle prediction can be combined with global speed planning and local speed planning to achieve a more extended time-domain prediction. Secondly, the driving cycle prediction only considers vehicle velocity prediction but does not consider ramps, road resistance coefficients. It can also be combined with the environment perception system in unmanned vehicles to achieve comprehensive prediction under complex road conditions.